\documentstyle[12pt]{article}

\begin{document}
\begin{titlepage}
\begin{flushleft}
Stockholm\\
USITP 96-12\\
September 1996\\
\end{flushleft}
\vspace{1cm}
\begin{center}
{\Large ON CHIRAL P-FORMS}\\
\vspace{15mm}
\vspace{5mm}
{\large Ingemar Bengtsson}\footnote{Email address: 
ingemar@vana.physto.se}\\
\vspace{5mm}
{\sl Fysikum\\
Stockholm University\\
Box 6730, S-113 85 Stockholm, Sweden}\\
\vspace{10mm}
{\large Astri Kleppe}\footnote{Email address:
astri@vana.physto.se}\\
{\sl Department of mathematical physics\\
Lund Institute of Technology\\
P.O. Box 118, S-221 00 Lund, Sweden}\\
\vspace{15mm}
{\bf Abstract}\\
\ \\
\end{center}
Some aspects of chiral p-forms, in particular the obstruction 
that makes it hard to define covariant Green functions, are 
discussed. It is shown that a proposed resolution involving 
an infinite set of gauge fields can be made mathematically 
rigourous in the classical case. We also give a brief 
demonstration of species doubling for chiral bosons.
\ \\
\end{titlepage}

\noindent {\bf 1. INTRODUCTION}

\vspace*{5mm}

\noindent What does it mean to say that a theory is "relativistic"?
 As a matter of fact two {\em a priori} different meanings of the
 term "relativistic invariance" are in use \cite{Sudarshan}. The
 first demands that there should be no preferred reference frame.
 Technically, the main requirement here is that there should exist
 a frame independent prescription for how to transform a description
 of a physical situation that uses a particular inertial frame to a
 description based on another inertial frame, and that the set of
 such transformations have to obey the multiplication laws of the
 Poincar\'{e} group. This is of course enough to make a search
 for scalar quantities meaningful. If the description of the
 system is based on the principle of least action, it is natural
 to demand also that the transformations should be implementable 
as canonical or unitary transformations. 

Another, and stronger, meaning of the term "relativistic
 invariance" requires that the description of the physical
 system should be "manifestly covariant", that is to say
 based entirely on finite dimensional tensors (and spinors).
 This demand is not only made for reasons of practical 
convenience, but also for a deeper philosophical reason: 
It ensures that the theory can be formulated in a way that
 is entirely independent of reference frames. 

Note that, throughout, we are concerned only with the special
 theory of relativity.

Now it is natural to ask: how much stronger than the first is
 the second definition of "relativistic invariance"? One might
 think that it is much stronger, or alternatively one might 
think that the two definitions are equivalent. In four 
spacetime dimensions, if we stick to the kind of models that
 can be described by field theories and restrict our attention
 to non-interacting systems, the latter answer can be shown
 to be correct by means of explicit constructions. Indeed it
 is known that to every irreducible representation of the 
Poincar\'{e} group (except the continuous spin representations,
 which we ignore) there corresponds a manifestly covariant 
free field theory whose solution space carries the 
representation in question \cite{Fronsdal}. Curiously, the 
situation in hypothetical spacetimes of higher dimensions is
 less clear. Marcus and Schwarz \cite {Marcus} made the 
remarkable claim that chiral p-forms \cite{Nahm} --- a kind
 of fields that can occur in space-times of twice odd dimension
 --- do not admit a manifestly covariant description, although
 they are relativistically invariant according to the first 
definition. Consequently the two possible meanings of 
``relativistic invariance'' would not be equivalent in such 
spacetimes. The direct relevance to physics of this 
counterexample is not obvious, although chiral p-forms do enter
 in the spectrum of type IIB superstrings, so that they 
occupy a small corner of a serious theory which may be of 
physical relevance. (A string that couples directly to a 
chiral p-form was constructed recently \cite{Schwarz}.) Be
 that as it may, our interest here is entirely in the questions
 of principle that the existence of this model raises. The 
principles are of physical relevance, and we wish to understand
 them.

A chiral p-form has a number of properties which are very 
peculiar. Except for the fact that its excitations obey 
Bose statistics, it shares many of the features of fermionic 
fields: it obeys a field equation which is linear in 
derivatives, it leads to species doubling if one attempts to 
define it on a lattice \cite{Tang}, and it gives rise to 
gravitational anomalies \cite{Witten}. A literature on the 
subject already exists --- and an even larger number of papers
 have been written on the simplest case, chiral bosons in 
$1 + 1$ dimensions, which from our point of view is a 
degenerate case since it leads to no difficulties with 
respect to covariance. The main reason why we intend to 
return to the subject in the present paper is simply that
 we believe that we can present some of the issues in a 
way that is somewhat more clear than that of previous 
treatments. Some of our observations are new, however.

In section 2 of this paper we try to explain the sense in which 
chiral p-forms violate manifest covariance. As we will emphasize,
 it is not enough to find a covariant action. One must also write
 down covariant Green functions, which is a harder task. We use 
the Hamiltonian formulation of the model to explain why it is hard,
 since it is enough to study the equal time restrictions for this
 purpose. The conclusion is that manifest covariance is violated 
in the same sense as manifest covariance is violated by 
non-covariant gauges (such as the Coulomb gauge) in 
electrodynamics. In section 3 we discuss a proposed solution 
of the problem which involves an infinite set of auxiliary 
fields \cite{McClain}, and we give a rigourous version of this 
proposal in the classical, 1 + 1 dimensional case. 
In section 4 we turn to species 
doubling. Concentrating on chiral bosons in two spacetime 
dimensions, we show - in a way that we find more direct than 
the original demonstration \cite{Tang} - precisely why species
 doubling occurs. Section 5 summarizes open problems.

\vspace*{1cm}

\noindent {\bf 2. MANIFEST COVARIANCE}

\vspace*{5mm} 

\noindent Let us assume that the space-time dimension $D$ is 
even, but twice odd, so that

\begin{equation} D = 2n = 2(p + 1) \ , \end{equation}

\noindent where $p$ is an even number. We also assume that we
 are in flat Minkowski spacetime, since nothing is gained by 
considering a more general situation. Then we define a p-form,
 that is to say a totally antisymmetric tensor field 
$A_{{\alpha}_1....{\alpha}_p}$ with $p$ indices, and we 
also introduce the notation

\begin{equation} A_{{\alpha}[p]} \equiv A_{{\alpha}_1....{\alpha}_p}
 \ . \end{equation}

\noindent Thus ${\alpha}[p]$ stands for an antisymmetric 
collection of $p$ indices. Further developments of this 
notation, such as how to handle index contractions, should
 be self evident. We will use it when it is convenient to do so.

Next we introduce the field strength

\begin{equation} F_{{\alpha}[n]} = 
n!\partial_{[{\alpha}_1}A_{{\alpha}_2...{\alpha}_n]} \ , 
\end{equation}

\noindent where the square brackets denote anti-symmetrization
 with weight one. There is an action for such objects that is
 a natural generalization of the Maxwell action, namely

\begin{equation} S[A] = - \frac{1}{2n!}\int d^{2n}x \ 
F_{{\alpha}[n]}F^{{\alpha}[n]} 
\ . \label{4} \end{equation}

\noindent This defines the dynamics of an ordinary p-form. 
Next we define the Hodge dual

\begin{equation} \star F_{{\alpha}[n]} = 
\frac{1}{n!\sqrt{-g}}g_{{\alpha}_1{\beta}_1}...g_{{\alpha}_n{\beta}_n}
{\epsilon}^{{\beta}[n]{\gamma}[n]}F_{{\gamma}[n]} \ . \end{equation}

\noindent Since $n$ is odd, and

\begin{equation} \star \star F_{{\alpha}[n]} = F_{{\alpha}[n]}
 \end{equation}

\noindent in Minkowski space, the eigenvectors of the 
$\star $-operator are real. Therefore the self-dual part 
of the field strength, which obeys

\begin{equation} \star F_{{\alpha}[n]} = F_{{\alpha}[n]} \ , 
\label{7} \end{equation}

\noindent is real. This is an important difference between 
twice odd and twice even dimensions. There is another difference
 that is even more relevant to us, and we will come to it soon.

The dynamics of a chiral p-form is defined by eq. (\ref{7}), 
which from now on we will adopt as our field equation. 
Solutions of this equation automatically obey the field 
equations derived from the action (\ref{4}). The simplest 
example of the self-duality condition occurs in $D = 1 + 1$ 
dimensions, where the potential is a scalar; 

\begin{equation} F_{\alpha} = \partial_{\alpha}{\phi} \ . 
\end{equation}

\noindent Then the self-duality condition (\ref{7}) becomes 
the equation for a chiral boson. Indeed, writing it out in 
components we find

\begin{equation} \star F_{\alpha} = F_{\alpha} \hspace*{1cm} 
\Leftrightarrow \hspace*{1cm} \dot{\phi} = \phi' \ , 
\end{equation}

\noindent where the dot and the prime denote time and space 
derivatives, respectively. It is often convenient to keep 
this example in mind when following the calculations for 
higher dimensions.

So far, everything is manifestly covariant in arbitrary $D$.
 The difficulty first appears when one attempts to derive the
 self-duality condition from an action. Actually it is not
 so easy to find an action, unless one first casts the 
equations in Hamiltonian form \cite{IB} \cite{Claudio}. 
Naturally, the Hamiltonian formulation itself violates 
manifest covariance in a sense. However, this happens in a
 perfectly controlled way, and there is therefore no reason 
why one should not use the Hamiltonian formulation to discuss 
the issue of manifest covariance \cite{Sudarshan}. Within 
the Hamiltonian formulation we will see the problem when we try
 to write down the Poisson brackets --- that is to say the 
equal time restriction of the commutator Green function. In 
the Hamiltonian as well as in the Lagrangian formulation, 
manifest covariance is violated when one inverts a certain 
``matrix'', to define respectively the Poisson brackets and 
the Green functions. A manifestly covariant Lagrangian
 is simply not enough. 

We begin, then, with the phase space version of the action for
 an ordinary p-form:

\begin{equation} S = \int d^{2n}x \ [\dot{A}_{a[p]}E_{a[p]} - 
\frac{1}{2p!}(E_{a[p]}E_{a[p]} + B_{a[p]}B_{a[p]}) -
p{\Lambda}_{a[p-1]t}\partial_{a_1}E_{a_1a[p-1]}] \ . \end{equation}

\noindent Here the ``electric'' field $E_{a[p]}$ is the 
momentum canonically conjugate to the p-form, and $B_{a[p]}$
 is the ``magnetic'' field. Explicitly,

\begin{equation} E_{a[p]} = F_{ta[p]} \hspace*{2cm} B_{a[p]} 
\equiv \frac{1}{n!}{\epsilon}_{a[p]b[n]}F_{b[n]} \ . \end{equation}

\noindent Also the time-space component $A_{ta[p-1]}$ enters the
 action in the guise of the Lagrange multiplier 
${\Lambda}_{a[p-1]}$, giving rise to a first class constraint,
 which is analogous to Gauss' law for the Maxwell theory. We 
wish to constrain this model further, so that only self-dual 
solutions occur. In canonical language, eq. (\ref{7}) becomes

\begin{equation} E_{a[p]} = B_{a[p]} \ . \label{10} \end{equation}

\noindent It is straightforward to amend the phase space action
 so that it gives rise to this constraint: from now on we consider 

\begin{equation} S = \int d^{2n}x \ [\dot{A}_{a[p]}E_{a[p]} - 
\frac{1}{2p!}(E_{a[p]}E_{a[p]} + B_{a[p]}B_{a[p]}) - 
{\Lambda}_{a[p]}(E_{a[p]} - B_{a[p]})] \ . \label{13} \end{equation}

\noindent The only change is in the Lagrange multiplier. The
 constraint is given by eq. (\ref{10}).

We now come to the second crucial difference between 
space-times whose dimension are twice odd or twice even. 
In the twice odd case, the self-duality constraint is 
partially second class (in Dirac's terminology \cite{Dirac})
, since

\begin{eqnarray} \{E_{a[p]}(x) - B_{a[p]}(x), E_{b[p]}(y) - 
B_{b[p]}(y)\} 
= \hspace*{3cm} \nonumber \\
\ \label{11} \\
\hspace*{15mm} 
= - (1 + (-1)^p){\epsilon}_{a[p]cb[p]}\partial_c{\delta}(x,y) = 
- 2{\epsilon}_{a[p]b[p]c}\partial_c{\delta}(x,y) \ . 
\nonumber \end{eqnarray}

\noindent By contrast, in the twice even case $p$ is odd, 
the right hand side vanishes, and all the constraints are 
first class, which
 means that a new gauge symmetry appears. There will then be
 no local degrees of freedom left in the model. Returning to 
the twice odd case, it is crucial to observe that the self-duality 
constraint contains both second and first class 
constraints. It contains second class constraints because the
 right hand side of eq. (\ref{11}) is non-zero, but it also 
contains first class constraints, because the divergence of 
the ``magnetic'' field is identically zero, so that eq. 
(\ref{10}) implies ``Gauss' law''. Gauss' law is first class 
and remains as a generator of gauge transformations.

In the twice odd case the Hamiltonian is non-vanishing also when
 the self-duality constraint is imposed (whereas the Hamiltonian
 is zero for self-dual configurations in twice even dimensions).
 We can summarize the above discussion by the weak equality

\begin{equation} H \approx \frac{1}{p!} \int d^{n+p}x 
\ B_{a[p]}B_{a[p]} \ , 
\label{12} \end{equation}

\noindent with the understanding that we have to take into 
account the constraints

\begin{equation} {\Phi}_{a[p]} \equiv E_{a[p]} - B_{a[p]} \approx 0 
\ , \label{16} \end{equation}

\noindent where $E_{a[p]}$ is the ``naive'' canonical momentum.

The next step, which has to be taken before we can say that we have
 defined a consistent Hamiltonian system, is to solve the second 
class constraints. To do this it is necessary to isolate the 
second class part of eq. (\ref{16}). There is no unique way to 
do this, but all ways entail a violation of manifest covariance.
 A possibility that springs immediately to mind is to introduce
 the transverse and longitudinal projection operators

\begin{equation} {\Pi}^T_{ab} \equiv {\delta}_{ab} - 
\frac{1}{\Delta}\partial_a\partial_b \hspace*{1cm} {\Pi}^L_{ab} 
\equiv
 \frac{1}{\Delta}\partial_a\partial_b \ , \end{equation}

\noindent where ${\Delta}$ is the Laplacian. Then we can decompose

\begin{eqnarray} A_{a[p]} &=& A^T_{a[p]} + A^L_{a[p]} \ , 
\nonumber \\
\ \\
A^T_{a[p]} &\equiv & {\Pi}^T_{a_1b_1}...
{\Pi}^T_{a_pb_p}A_{b[p]} \ , 
\nonumber \end{eqnarray}

\noindent and similarly for all other objects. We then see that
 the transverse part of the constraint is second class, and the
 longitudinal part first class:

\begin{eqnarray} \{{\Phi}^T_{a[p]}(x), {\Phi}^T_{b[p]}(y)\} &=& 
-2{\epsilon}_{a[p]b[p]c}\partial_c{\delta}(x,y) \nonumber \\
\ \\
\{{\Phi}^L_{a[p]}(x), {\Phi}^L_{b[p]}(y)\} &=& 0 \ . \nonumber 
\end{eqnarray}

\noindent We can now solve the second class constraints for the 
transverse ``electric'' field, and compute the resulting Dirac 
brackets for the transverse vector potential. They are

\begin{equation} \{A^T_{a[p]}(x), A^T_{b[p]}(y)\} = 
\frac{1}{2p!} {\epsilon}_{a[p]b[p]c} 
\frac{1}{\Delta}\partial_c{\delta}(x,y) \ . \label{17} \end{equation}

\noindent Now we have arrived at a consistent Hamiltonian system,
 described by the Hamiltonian (\ref{12}), the Poisson brackets 
(\ref{17}), the naive Poisson brackets for the longitudinal 
parts of the original dynamical variables, and the first class 
constraints ${\Phi}^L_{a[p]} \approx 0$. No gauge fixing has 
been performed, and yet our formul\ae \ are non-local and 
they violate manifest covariance in exactly the same sense that 
the Coulomb gauge in electrodynamics violates manifest 
covariance. The violation is 
caused by the non-covariant and non-local Poisson brackets. 

When $D = 1 + 1$ the transverse part of the potential is missing.
 Therefore --- as noted by Marcus and Schwarz \cite{Marcus} --- 
there is no problem with covariance in this case. The only 
remaining complication is that the boundary conditions have to be 
chosen such that the Dirac brackets are well defined.

Although no gauge has been fixed, there was an element of choice
 in the above discussion. The set of constraints ${\Phi}_{a[p]}$
 that we had at the outset can be decomposed into first and second
 class constraints in other ways. For instance, if we let the 
index $i$ range from $1$ to $D-2$, and use $z$ to denote the 
remaining spatial coordinate, we find a subset of the constraints
 which obeys

\begin{equation} \{{\Phi}_{i[p]}(x), {\Phi}_{j[p]}(y)\} = 
-2{\epsilon}_{i[p]j[p]}\partial_z{\delta}(x,y) \ . \end{equation}

\noindent This subset of the constraints can be declared to be 
second class, and once the Dirac brackets have been computed one 
sees that the remaining constraints are first class. We arrive at
 a formulation which differs from the first, but again manifest 
covariance is violated --- this time in exactly the same sense 
that the axial gauge in electrodynamics violates manifest 
covariance. There is simply no covariant way to define the 
symplectic structure that we are studying, and by implication the
 same statement is true for the Green functions. 

This violation of manifest covariance is rather mild, however. 
The gauge invariance that is also present in the model selects a
 set of gauge invariant observables, namely the ``magnetic'' 
fields $B_{a[p]}$. The Dirac bracket

\begin{equation} \{B_{a[p]}(x), B_{b[p]}(y)\} = - {\epsilon}_
{a[p]b[p]c}\partial_c{\delta}(x,y) \end{equation}

\noindent is perfectly local, and the energy-momentum densities
 --- which can be expressed entirely in terms of $B_{a[p]}$ ---
 transforms this field in a covariant way. (This point was 
stressed by Henneaux and Teitelboim \cite{Claudio}.) This is
 again similar to electrodynamics in the Coulomb gauge.

\vspace*{1cm}

\noindent {\bf 3. A COVARIANT ACTION}

\vspace*{1cm} 

\noindent An obvious Lagrangian for chiral p-forms is obtained by 
performing a Legendre transformation of the Hamiltonian that we 
have studied. To do this it is convenient to split the Lagrange 
multiplier ${\Lambda}_{a[p]}$ into transverse and longitudinal 
parts, and use the longitudinal part as the time-space component
 of the potential. In this way we can rewrite the phase space 
action (\ref{13}) as

\begin{equation} S = \frac{1}{p!} \int d^{2n}x \ [F_{ta[p]}E_{a[p]} - 
\frac{1}{2}(E_{a[p]}E_{a[p]} + B_{a[p]}B_{a[p]}) - 
p!{\Lambda}^T_{a[p]}(E_{a[p]} - B_{a[p]})] \ . \end{equation}

\noindent If we integrate out first $E_{a[p]}$ and then 
${\Lambda}^T_{a[p]}$ we obtain a Lagrangian which is, however,
 not manifestly covariant. Thus the naive attempt to derive a
 manifestly covariant Lagrangian does not work. 

Is it possible to do better? Indeed Siegel \cite{Siegel} quickly
 found a covariant Lagrangian, which consists of the ordinary 
action for a p-form with a certain cubic term added. A correct
 analysis \cite{IB} shows that Siegel's covariant action leads
 to precisely the same Hamiltonian formulation as the one we
 have studied. Therefore Siegel's action is covariant, but it
 can not be used to derive covariant Green functions. At this
 point then, the claim by Marcus and Schwarz still stands. 
Incidentally, this is quite similar to the situation for the
 Green-Schwarz superstring \cite{Martin}, although the latter
 is not a field theory action, and moreover it has at least 
one alternative, fully covariant description, namely the NSR formalism.

However, some years later a manifestly covariant formulation 
of chiral p-forms was in fact presented \cite{McClain}. 
(Actually the original reference treats only the two 
dimensional case, but the generalisation to higher dimensions
 is straightforward \cite{Devecchi}.) An awkward feature of 
this formulation is that it requires an infinite set of 
auxiliary fields. The phase space action is

\begin{equation} S = \sum_{n=0}^{\infty} \int d^6x \ 
[\dot{A}_{ab}^{(n)}E^{(n)}_{ab} - \frac{(-1)^n}{4}
(E_{ab}^{(n)}E_{ab}^{(n)} + B_{ab}^{(n)}B_{ab}^{(n)}) + 
{\Lambda}_{ab}^{(n+1)}{\Psi}_{ab}^{(n+1)} - 2
 A_{ta}^{(n)}{\cal G}_a^{(n)}] , \end{equation}

\noindent where the action depends on an infinite set of 
fields and Lagrange multipliers labelled by the index $n$,
 and we have specified $D = 5+1$ for definiteness. The 
constraints are

\begin{equation} {\cal G}_a^{(n)} = \partial_bE_{ab}^{(n)} 
\approx 0 \end{equation}

\begin{equation} {\Psi}^{(n+1)}_{ab} = E^{(n)}_{ab} - 
B^{(n)}_{ab} + E^{(n+1)}_{ab} + B^{(n+1)}_{ab} \approx 0
 \ . \end{equation}

\noindent Note that this time the constraints ${\Psi}^{(n)}
 \approx 0$ do not imply that Gauss' law holds, and therefore
 the latter must be explicitly added. It is easy to check 
that the Hamiltonian is weakly gauge invariant and that all
 the constraints are first class. The corresponding 
Lagrangian is manifestly covariant:

\begin{equation} S = \sum_{n=0}^{\infty} \int d^6x \ 
\frac{(- 1)^n}{3}\left( - \frac{1}{4}F_{{\alpha}{\beta}
{\gamma}}^{(n)}F^{{\alpha}{\beta}{\gamma}}_{(n)} - 
{\Lambda}^{{\alpha}{\beta}{\gamma}}_{(n+1)}(F_{{\alpha}
{\beta}{\gamma}}^{(n)} - F_{{\alpha}{\beta}{\gamma}}^{(n+1)})
 + {\Lambda}^{{\alpha}{\beta}{\gamma}}_{(n+1)}{\Lambda}_
{{\alpha}{\beta}{\gamma}}^{(n+2)}\right) \ . \end{equation}

\noindent Here we have set

\begin{equation} {\Lambda}_{ab} \equiv {\Lambda}_{tab}
 \end{equation}

\noindent and the Lagrange multiplier fields are alternately
 self-dual and anti-self-dual;

\begin{equation} \star {\Lambda}_{{\alpha}{\beta}{\gamma}}^{(n)}
 = (- 1)^n {\Lambda}_{{\alpha}{\beta}{\gamma}}^{(n)}
 \ . \end{equation}

\noindent Covariant gauge fixing of this action is --- at 
least formally --- straightforward \cite{Devecchi}, so 
that covariant Green functions can be written down. 

Of course we have yet to show that the action really
 describes a single chiral p-form. A naive argument which
 suggests that it does goes as follows: first use Gauss' 
Law to set the longitudinal degrees of freedom to zero. Then
 fix the remaining gauge freedom by the condition

\begin{equation} E^{(n)}_{ab} + B^{(n)}_{ab} = 0 \ , 
\hspace*{2cm} n > 0 \ . \end{equation}

\noindent Now all fields with $n > 0$ vanish as a consequence
 of the constraints, and the fields with $n = 0$ obey the same
 equations that we studied before, hence they do indeed describe
 a single chiral p-form. However, because of the occurence of
 an infinite set of fields in the covariant action it is 
necessary to scrutinize this naive argument carefully to see 
that no inconsistencies arise. We remark in passing that the
 same action but with only a finite set of $N$ fields describes
 either one self-dual and one anti-self-dual p-form, or a pair
 of self-dual p-forms one of which contributes with a negative
 sign to the total energy, depending on whether $N$ is odd or
 even. Hence the infinite set can not be avoided. Incidentally
 this is similar to some covariant modifications of the Green-
Schwarz type of action for superparticles \cite{Kallosh}.

To study the consistency of the covariant formulation we restrict
 ourselves to the original McClain-Wu-Yu action in $1+1$ 
dimensions \cite{McClain};

\begin{equation} S = \sum_{n=0}^{\infty} \int d^2x \ 
[\dot{\varphi}_{(n)}{\pi}_{(n)} - \frac{(- 1)^n}{2}({\pi}_
{(n)}^2 + \varphi'^{2}_{(n)}) - {\Lambda}_{(n+1)}
{\Psi}_{(n+1)}] \ , \end{equation}

\noindent where $\varphi_{(n)}$ are an infinite collection
of scalar fields and $\pi_{(n)}$ their momenta, and the constraints are

\begin{equation} {\Psi}_{(n+1)} = {\pi}_{(n)} - 
\varphi'_{(n)} + {\pi}_{(n+1)} + 
\varphi'_{(n+1)} \approx 0 \ . \end{equation}

\noindent It is easy to give an example of a formal calculation 
which leads to a contradiction \cite{Clovis}. Thus, define

\begin{equation} {\Pi} \equiv \sum_{n = 0}^{\infty} {\pi}_{(n)} 
\ . \end{equation}

\noindent This expression is formally gauge invariant. A formal 
addition of the constraints gives

\begin{equation} \label{etti}
{\Pi} = \frac{1}{2}(\pi_{(0)} + 
\varphi'_{(0)}) + \frac{1}{2}\sum_{n = 0}^{\infty} 
{\Psi}_{(n)} \approx \frac{1}{2}(\pi_{(0)} + 
\varphi'_{(0)}) \equiv I_0 \ . \end{equation}

\noindent There should then be two equivalent ways of expressing 
our single gauge invariant degree of freedom. However, 

\begin{equation} \{I_0(x), I_0(y)\} = \frac{1}{2}\partial_x
{\delta}(x,y) 
\ , \end{equation}

\noindent which is weakly non-zero, while the Poisson bracket 
$\{{\Pi}(x), {\Pi}(y)\} = 0$. This contradicts the general theorem 
that 

\begin{equation} \tilde{F} \approx F \hspace{5mm} \& 
\hspace{5mm} \tilde{G} \approx G \hspace{1cm} \Rightarrow 
\hspace{1cm} \{\tilde{F}, \tilde{G}\} \approx \{F, G\} \ , 
\end{equation}

\noindent so that the naive approach is clearly wrong. 

If one tries to do the calculation in detail, one sees that the 
problem occurs because one encounters the infinite sum

\begin{equation} \label{tvaa}
A_{total} = A(1 - 1 + 1 - 1 + 1 - 1 + \ ... \ ) 
\ . \end{equation}

\noindent A similar sum appears when one makes a naive 
attempt to calculate the 
gravitational anomaly in the quantum version of the model. One 
can of course look for a regularization procedure that permits 
one to set this sum equal to the answer one desires, but we 
prefer to formulate the theory in such a way that the sum never 
occurs, and we will now proceed to do this for the classical 
theory.

To avoid the many fallacies that appear with an unlimited number 
of variables we clearly need to specify what kinds of functions 
$\varphi_{(n)}$, $\pi_{(n)}$ that are allowed in the phase space 
of the model. In order that

\begin{equation}
{\cal{H}}=\frac{1}{2} \int d^{2}x \displaystyle\sum_{n=0}^{\infty} 
(-1)^{n}(\pi^{2}_{(n)}+\varphi'^{2}_{(n)}) =\frac{1}{4} \int d^{2}x 
\displaystyle\sum_{n=0}^{\infty}(-1)^{n} 
[(\pi_{(n)}+\varphi'_{(n)})^{2}+
(\pi_{(n)}-\varphi'_{(n)})^{2}] \end{equation} 
                          
\noindent be finite, we demandthat the fields and momenta are 
limited from above, in the sense that there exist some continuous 
and differentiable functions $f$ and $h$ such that

\begin{equation}
\int d^{2}x [f^{2}(x)+h^{2}(x)]\nonumber
\end{equation}

\noindent is finite and well-defined, and

\begin{equation}\label{cond}
|\pi_{(n)}(x)+\varphi'_{(n)}(x)|\leq{\hspace{2mm}}\frac{f(x)}{n},
{\hspace{4mm}}n>N_{0}\nonumber \end{equation}

\begin{equation}\label{cind}
|\pi_{(n)}(x)-\varphi'_{(n)}(x)|\leq{\hspace{2mm}}\frac{h(x)}{n},
{\hspace{4mm}}n>N_{0}\nonumber \end{equation}

\noindent for all $x=|\vec{x}|$.
If this is satisfied, the Hamiltonian is well-defined and finite, since 
$I_{0}$ is finite and
\begin{equation}
\int d^{2}x \displaystyle\sum_{n=1}^{\infty} (-1)^{n}(\pi^{2}_{(n)}+
\varphi'^{2}_{(n)})\leq\nonumber\\
\frac{1}{2}
\int d^{2}x \displaystyle\sum_{n=1}^{\infty}(-1)^{n}\frac{1}{n^{2}}
(h^{2}+f^{2}) = \frac{\pi^{2}}{24}\int d^{2}x(h^{2}+f^{2}) .
\end{equation}

The convergence property should be gauge invariant 
as well as time translation invariant. Consider the gauge 
transformations generated by the first class constraints,

\begin{eqnarray}
\delta \varphi_{(n)}& = &\{\varphi_{(n)},\int \epsilon_{(n)}
\Psi_{(n)}\}=\epsilon_{(n+1)}+\epsilon_{(n)}\nonumber\\
    \delta \pi_{(n)}& = &\{\pi_{(n)},\int \epsilon_{(n)}\Psi_{(n)}\}
=\epsilon'_{(n)}-\epsilon'_{(n+1)}\end{eqnarray}

\noindent and denote the transformed momenta and fields by

\begin{eqnarray}
\tilde{\pi}_{(n)}=\pi_{(n)}+\delta\pi_{(n)}{\hspace{4mm}}{\rm{and}}
{\hspace{4mm}}
\tilde{\varphi}_{(n)}=\varphi_{(n)}+\delta\varphi_{(n)} .\nonumber
\end{eqnarray}

\noindent The gauge transformed series 
\begin{eqnarray}
\frac{1}{2}\displaystyle\sum_{n=0}^{\infty}(-1)^{n}
(\tilde{\pi}^{2}_{(n)}+\tilde{\varphi}'^{2}_{(n)}) \nonumber
\end{eqnarray}

\noindent is obviously convergent if we demand that the gauge 
parameters $\epsilon_{(n)}$ are such that

\begin{equation}\label{raxi}
\epsilon'_{(n)} =\epsilon'_{(n)}(x) \leq \frac{k(x)}{n},{\hspace{5mm}}
n>N_{0}
\end{equation}

\noindent for some suitable function $k(x)$.

By introducing the restriction ${{(\ref{raxi})}}$ on the gauge 
parameters, the inconsistency that is brought about by the naive 
approach in ${{(\ref{etti})}}$ - ${{(\ref{tvaa})}}$ is avoided, 
just as was desired.

We next inspect the convergence property under time translation by 
expressing the fields as solutions to the Klein-Gordon equation, viz.

\begin{equation}
\varphi_{(n)}=B_{(n)}(x+t)+G_{(n)}(x-t)
\end{equation}

\noindent where $B$ and $G$ are continuous, differentiable functions 
with derivatives

\begin{eqnarray}
\frac{ \partial}{\partial y}B_{(n)}(y)=b_{(n)}(y){\hspace{5mm}}
{\rm{and}}{\hspace{5mm}}
\frac{ \partial}{\partial y}G_{(n)}(y)=g_{(n)}(y) .\nonumber
\end{eqnarray}

\noindent Since $\dot{\varphi}_{(n)}=\pi_{(n)}$, this implies that
$\pi_{(n)}+\varphi'_{(n)}=2b_{(n)}(x+t)$ and
$\pi_{(n)}-\varphi'_{(n)}=-2g_{(n)}(x+t)$.
If ${{(\ref{cond})}}$ and ${{(\ref{cind})}}$ are satisfied at 
the initial time $t=0$, we get that
\begin{eqnarray}\label{oxxo}
|\pi_{(n)}(x)+\varphi'_{(n)}(x)|=|2b_{(n)}(x) |\leq \frac{f(x)}{n}
\end{eqnarray}

\noindent for all $x$. This however implies that

\begin{eqnarray}\label{oxxi}
|\pi_{(n)}(x,t)+\varphi'_{(n)}(x,t)|=|2b_{(n)}(x+t)| \leq 
\frac{f(x+t)}{n}\end{eqnarray}

\noindent for all $t$ as well as for all $x$, because we can 
always rewrite the argument as\\ $x+t=x'$. The same goes for

\begin{eqnarray}
|\pi_{(n)}(x,t)-\varphi'_{(n)}(x,t)|=|2g_{(n)}(x-t)| 
\leq \frac{h(x-t)}{n} .\end{eqnarray}

\noindent So in conclusion, if the Hamiltonian is initially 
well-defined and finite, this also holds at any later times.

We finally investigate the form of the Lagrange multipliers.
If we choose the gauge fixing condition

\begin{eqnarray}
\pi_{(n)}+\varphi'_{(n)}=0, {\hspace{10mm}}n>0\nonumber
\end{eqnarray}

\noindent it follows that

\begin{eqnarray}
\pi_{(n)}-\varphi'_{(n)}=0, {\hspace{3mm}}{\rm{all}}{\hspace{2mm}}n,
\nonumber\end{eqnarray}

\noindent and only one degree of freedom remains, represented e.g as 
$\varphi'_{(0)}$
or $\pi_{(0)}+\varphi'_{(0)}$, with the corresponding Hamiltonian

\begin{eqnarray}
{\cal{H}}=\frac{1}{4}(\pi_{(0)}+\varphi'_{(0)})^{2} .\nonumber
\end{eqnarray}

\noindent Now, to guarantee that the Hamiltonian satisfies the weak 
equality

\begin{equation}\label{ling}
{\cal{H}}=\frac{1}{2} \int d^{2}x \displaystyle\sum_{n=0}^{\infty} 
(-1)^{n}(\pi^{2}_{(n)}+\varphi'^{2}_{(n)}) = \frac{1}{4}(\pi_{(0)}+
\varphi'_{(0)})^{2}-\displaystyle\sum_{n=1}^{\infty} \Lambda_{(n)}
\Psi_{(n)}\approx  \frac{1}{4}(\pi_{(0)}+\varphi'_{(0)})^{2}
\end{equation}                     
      
\noindent the gauge parameters must take the form

\begin{equation}\label{posi}
\Lambda_{(n)}=\frac{1}{4}(-1)^{n}[\pi_{(n)}+\varphi'_{(n)}-
\pi_{(n-1)}+\varphi'_{(n-1)}] .
\end{equation}

\noindent Since

\begin{eqnarray}
|\pi_{(n)}+\varphi'_{(n)}-\pi_{(n-1)}+\varphi'_{(n-1)}|\leq
|\pi_{(n)}+\varphi'_{(n)}|+|\varphi'_{(n-1)}-\pi_{(n-1)}|\leq 
\frac{f(x)}{n}+ \frac{h(x)}{n} ,\nonumber
\end{eqnarray}

\noindent $|\Lambda_{(n)}|$ is clearly limited by some quantity 
$\sim 1/n$, and therefore the $\Lambda_{(n)}$ belong to the set of 
gauge parameters defined in ${{(\ref{raxi})}}$.

In conclusion, if we demand that the fields and momenta satisfy the 
conditions ${{(\ref{cond})}}$ and ${{(\ref{cind})}}$, 
and that the gauge parameters 
satisfy ${{(\ref{raxi})}}$, we obtain a consistent
covariant formulation of the action.

\vspace*{1cm}

\noindent {\bf 4. SPECIES DOUBLING}

\vspace*{1cm}

\noindent Since chiral p-forms give rise to gravitational
 anomalies \cite{Witten}, it is not surprising that they 
also exhibit species doubling if one attempts to define 
them on a lattice. Here we will present a direct demonstration
 of this fact for chiral bosons in two spacetime dimensions. 
It seems worthwhile to do this here, since the only demonstration
 in the literature \cite{Tang} makes use of an action that does
 not describe a single chiral boson, so that the argument 
becomes quite indirect. 

To begin, a Lagrangian that describes an ordinary scalar boson
 on a discretized space (with time kept continuous) is

\begin{equation}\label{pong}
{\cal{L}}=\frac{a}{2}\displaystyle\sum_{n}(\dot{\varphi}_
{n}^{2}-(\frac{\Delta\varphi_{(n)}}{a})^{2}) ,
\end{equation}
where $a$ is the lattice spacing and $n$ denotes the lattice
 sites, running over integer values
$1 \leq n \leq N $ where $N$ is an integer. To discuss the 
most general discretization possible it is convenient to 
introduce the Fourier transform

\begin{equation}
\Delta\varphi_{(n)}=\frac{1}{n}\displaystyle\sum_{k}
\Delta\varphi_{(k)}e^{-2\pi ikn/N} .
\end{equation}
We assume that for any $\Delta\varphi_{(n)}$, we can express 
$\Delta\varphi_{(k)}$ as 
\begin{equation}\label{efff}
\Delta\varphi_{(k)}=-if(k)\varphi_{(k)}
\end{equation}

\noindent where $f(k)$ is some periodic function of $k$. 
This is then the most general discretization that we will 
consider. In order to ensure that the theory be local on 
the lattice, $f(k)$ must be continuous.

Simple examples of discretizations are the symmetrical 
discretization

\begin{equation}\label{symm}
\Delta\varphi_{(n)}=\frac{1}{2}(\varphi_{(n+1)}-\varphi_{(n-1)}) 
\hspace*{8mm} \Leftrightarrow \hspace*{8mm} f(k) =
 \sin\frac{2\pi k}{n} \ ,
\end{equation}

\noindent and the asymmetrical discretization

\begin{equation}\label{aymm}
\Delta\varphi_{(n)} = \varphi_{(n+1)}-\varphi_{(n)} \hspace*{8mm}
 \Leftrightarrow \hspace*{8mm} f(k) = 2e^{-i\pi k/N}\sin
\frac{\pi k}{n} \ . 
\end{equation}

\noindent As is well known (if not the reader will be 
reminded soon), the symmetrical discretization leads to 
species doubling in the continuum limit, while the asymmetrical
 one does not.

The Hamiltonian corresponding to ${{(\ref{pong})}}$ is

\begin{eqnarray}
{\cal{H}}&=&\frac{a}{2}\displaystyle\sum_{n}(
\dot{\varphi}_{(n)}^{2}+(\frac{\Delta\varphi_{(n)}}{a})^{2})=
\nonumber\\
\ \\
         &=&\frac{a}{2N}\displaystyle\sum_{k}(\dot{\varphi}_{(k)}
\dot{\varphi}_{-k}-\frac{1}{a^{2}}\varphi_{(k)}\varphi_{-k}f(k)f(-k))
 \ . \nonumber
\end{eqnarray}

\noindent In order to ensure that ${\cal{H}}$ be real and positive,
 we have to require that 

\begin{equation} f(-k) = - f^{*}(k) \ , \label{40} \end{equation}

\noindent which implies that

\begin{equation}
{\cal{H}}= \frac{a}{2N}\displaystyle\sum_{k}(\dot{\varphi}_{(k)}
\dot{\varphi}_{-k}
+\frac{1}{a^{2}}\varphi_{(k)}\varphi_{-k}f(k)f^{*}k))>0 \ ,
\end{equation}

\noindent as desired.

We now add the self-duality requirement. This means that we 
impose the condition

\begin{equation} \dot{\varphi}_n = \frac{1}{a}{\Delta}
{\varphi}_n \hspace*{8mm} \Leftrightarrow \hspace*{8mm} 
\dot{\varphi}_k = - \frac{i}{a} f(k) {\varphi}_k \ . \end{equation}

\noindent However, the equation of motion that follows from our
 Lagrangian is, in momentum space,

\begin{equation}
a^{2}\ddot{\varphi}_{(k)} - \varphi_{(k)}f(k)f(-k) = 0 \ .
\end{equation}

\noindent Therefore the self-duality condition is consistent
 with the equations of motion if and only if

\begin{equation} f(- k) = - f(k) \ . \end{equation}

\noindent If we refer back to eq. (\ref{40}) we see that the 
self-duality requirement forces $f(k)$ to be a real and odd 
function. Since it is also periodic, it has to have an even 
number of zeroes. In particular, the asymmetrical discretization 
given above is not allowed by self-duality.

Now the point is that if we go to the continuum limit, each 
zero of the dispersion relation will contribute one degree of 
freedom to the continuum model, and the chirality of this degree 
of freedom will depend on the slope of $f(k)$ at its zero (see 
ref. \cite{Kogut}). For us, this implies that the continuum 
model will have an equal number of self-dual and anti-self dual 
degrees of freedom. This is species doubling.

\vspace*{1cm}

\noindent {\bf 5. OPEN QUESTIONS}

\vspace*{1cm} 

\noindent In section 2 of this paper we tried to make a point 
which has not been made very clear in the literature, namely 
that the obstruction to manifest covariance usually is most 
severe for the Green functions (and by implication for the 
Poisson brackets). A manifestly covariant Lagrangian may well 
exist in a model which lacks manifestly covariant Green 
functions. In section 3 we discussed a proposed 
resolution of the problem for chiral p-forms, due to McClain, 
Wu and Yu. Their proposal involves an infinite set of 
gauge fields, and we drew attention to some difficulties that 
must be resolved. In a situation with an infinite number of fields,
a straightforward naive treatment leads to expressions
like $A_{total} = A(1 - 1 + 1 - 1 + 1 - 1 + \ ... \ ) $, where
the sum depends upon the regularization procedure. This can 
however be avoided by using a more rigorous formulation.
 In the classical theory we showed that 
a rigourous formulation can indeed be given, provided 
that one restricts the phase space as well as the Lagrange 
multipliers that are allowed to smear the constraints suitably. 
In effect, one has to make a careful distinction between proper 
and improper constraints. We discussed only the case $D = 2$, 
but the generalization to higher dimensions should not offer 
any difficulties. However, we have left the definition of the 
quantum theory as an open question.

In section 4 we showed directly how species doubling occurs 
on the lattice in $1+1$ dimensions. 
It would be of some interest to consider the general case, and 
to bring it up to the level of rigour that exists for chiral 
fermions \cite{Holger}, but we have not done so.

It is clear that there is a close connection between species 
doubling and the gravitational anomaly. Alvarez-Gaum\'{e} and 
Witten \cite{Witten} argue that the latter is also closely 
connected to the lack of a manifestly covariant description. 
If the McClain-Wu-Yu proposal can be shown to work 
for the quantum theory with coupling to external fields 
included, then this argument will have to be scrutinized.

There are many interesting aspects of chiral p-forms that we
 have not touched at all. For instance, it has been suggested 
\cite{John} that the difficulties with manifest covariance that
 we have encountered are akin to those that stand in the way
 of electrodynamics coupled to both electric and magnetic charges. 
The question whether an alternative formulation in terms of 
fermionic fields exists (in any twice odd dimension) is another 
issue that ought to be looked into. Finally, a thoughtful comparison 
of chiral p-forms with the chiral formulation of Einstein's 
equations due to Ashtekar \cite{Ashtekar} (in the twice even 
dimension $D = 4$) could give rise to valuable insights.

\end{document}